\documentclass[11pt]{article}
\usepackage{moriond,epsfig}

\def\Journal#1#2#3#4{{#1} {\bf #2}, #3 (#4)}

\def\NPB{{\em Nucl. Phys.} B}
\def\PLB{{\em Phys. Lett.}  B}

\def\ZPC{{\em Z. Phys.} C}

\def\be{\begin{equation}}
\def\ee{\end{equation}}
\def\bea{\begin{eqnarray}}
\def\eea{\end{eqnarray}}

\begin{document}
\vspace*{4cm}
\title{PRODUCTION OF SECONDARIES IN pp COLLISIONS AT RHIC ENERGIES}

\author{G.H. Arakelyan, C. Merino, C. Pajares, and
Yu.M. Shabelski}

\address{Departamento de F\'\i sica de Part\'\i culas, Facultade de 
F\'\i sica, \\ 
and Instituto Galego de F\'\i sica de Altas Enerx\'\i as (IGFAE), \\ 
Universidade de Santiago de Compostela, Galiza, Spain}

\maketitle\abstracts{
The midrapidity inclusive densities of different secondaries are calculated 
in the framework of the Quark--Gluon String Model. The transfer of baryon 
number in rapidity space due to the gluon string junction propagation leads 
to a significant effect on the net baryon production. The numerical results 
are in reasonable agreement with RHIC experimental data.}

%\section{Guidelines}
%\subsection{Producing the Hard Copy}\label{subsec:prod}

The Quark--Gluon String Model (QGSM) and the Dual Parton Model are based on 
the Dual Topological Unitarization and describe quantitatively many features 
of high energy production processes. The model parameters were fixed 
\cite{KTM,KaPi,CaTran,KTMS,Sh,Sh1} by comparison of the theoretical results 
with experimental data. In the present paper we compare the QGSM predictions 
with the experimental data~\cite{abe} on midrapidity yields $dn/dy$
($\vert y \vert < 0.5$) for different secondaries produced in $pp$ 
collisions at RHIC energy ($\sqrt{s} = 200.$ GeV).

High energy interactions are considered in the QGSM as taking place via the 
exchange of one or several Pomerons, all elastic and inelastic processes 
resulting from cutting through or between Pomerons~\cite{AGK}. Inclusive 
spectra of hadrons are related to the corresponding fragmentation functions of 
quarks and diquarks, which are constructed using the Reggeon counting rules 
\cite{Kai}.

At very high energies and in the midrapidity region all fragmentation 
functions, which are usually written \cite{Kai} as 
$G^h_q(z) = a_h (1-z)^{\beta}$ ($z$ is the fraction of a quark or diquark
momentum carried out by the secondary hadron), are constants 
\begin{equation}
G_q^h(z) = a_h \ , 
\end{equation}
and they consistently lead~\cite{AKM} to
%%%%and they consistently \cite{AKM} lead to
\begin{equation}
\frac{dn}{dy}\ \sim \ g_h \cdot (s/s_0)^{\alpha_P(0) - 1}
\sim a^2_h \cdot (s/s_0)^{\alpha_P(0) - 1} \,.
\end{equation}
This corresponds to the one-Reggeon exchange diagram in Fig.~1a, which is 
the only diagram contributing to the inclusive density in the central region 
(AGK theorem \cite{AGK}). The main contribution at high energies comes from 
the case when both Reggeons in Fig.~1a are Pomerons. The intercept of the 
supercritical Pomeron $\alpha_P(0) = 1 + \Delta$, $\Delta = 0.139$ \cite{Sh}, 
is used in the numerical calculations. 

\begin{figure}[htb]
\centering
\vspace{-3.5cm}
\includegraphics[width=.55\hsize]{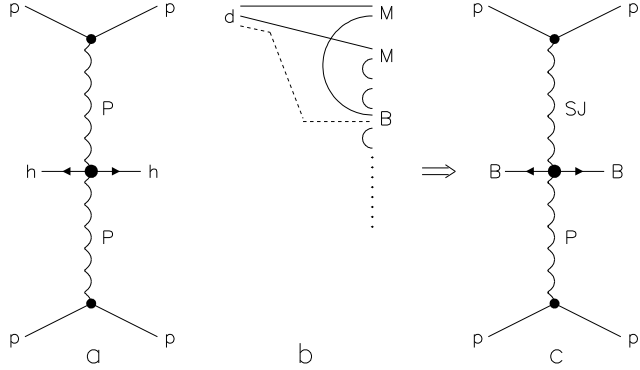}
\vskip -0.1cm
\caption{\footnotesize
One-Pomeron-pole diagram determining secondary hadron $h$ production (a).
String junction (shown by dashed line) diffusion (b) that leads to asymmetry 
in baryon/antibaryon production in the central region, and}
\vspace{-0.125cm} 
{\footnotesize \hspace{-10.2cm}the corresponding 
Reggeon diagram (c).}
\end{figure}

The diagram in Fig.~1a predicts equal inclusive yields for each particle and 
its antiparticle at very high energies. That is true for meson production,
but at RHIC energies a numerically small difference in the positive and
negative meson productions can exist. In the string models baryons are 
considered as configurations consisting of three connected strings (related 
to three valence quarks) called string junction  (SJ) \cite{IOT,RV}. In the 
processes of secondary production the SJ diffusion in rapidity space leads to 
significant differences in the yields of baryons and antibaryons in the 
midrapidity region even at very high 
energies \cite{Sh1,Bopp,ACKS,SJ1,AMS,AMPS,Olga,SJ2,SJ3}.

There exist \cite{ACKS} three different possibilities to obtain
the net baryon charge. The first one is the fragmentation of the diquark 
giving rise to a leading baryon. A second possibility is to produce 
a leading meson in the first break-up of the string and a baryon in the 
subsequent break-up \cite{Kai,22r}. In these two cases the baryon number 
transfer is possible only for short distances in rapidity. In the third 
case, which takes place with rather small relative probability, both initial 
valence quarks recombine with sea antiquarks into mesons and a secondary 
baryon is formed by the SJ together with three sea quarks.

An example is shown in  Fig.~1b where both valence quarks of the incident 
diquark annihilate with sea antiquarks into mesons $M$ and the baryon $B$ is 
produced rather far from the beam in rapidity space. The third valence quark 
of the incident baryon (not shown in Fig.~1b) independently fragments into a 
meson system. This process can be described by the Reggeon diagram shown in  
Fig.~1c which results in some corrections to the spectra of secondary
baryons. As $\alpha_{SJ}$ is close to unity (we use $\alpha_{SJ} = 0.9$ 
\cite{SJ1}) the difference of the baryon and antibaryon yields in 
midrapidity region will vanish very slowly when the energy increases.

The normalization constants in Eq.~(2) for pion production, $a_{\pi}$, kaon 
production, $a_K$, $B\bar{B}$ pair production, $a_{\bar{N}}$, and baryon 
production due to SJ diffusion, $a_N$, were determined  \cite{KTM,KaPi,Sh}
from the experimental data at fixed target energies. Their values are : 
\begin{equation}
a_{\pi} = 0.67\,,\; a_k = 0.21\,,\;a_{\bar{N}} =0.18\,,\;a_N =1.29
\,.
\end{equation}
The values of these 
parameters have not been modified for the present calculations, while the 
the corresponding values for hyperons have been calculated by 
simple quark combinatorics~\cite{AnSh,CS}. For sea quarks we have
\begin{equation}
p:n:\Lambda + \Sigma:\Xi^0:\Xi^-:\Omega = 4L^3:4L^3:12L^2S:3LS^2:3LS^2:S^3
\,.
\end{equation} 
The strangeness suppression factor is given by the ratio $\lambda = S/L$, and 
$2L+S = 1$. Usually, in soft processes $\lambda$ is assumed to have a value 
$\lambda = 0.2$-$0.35$. Inside this range $\lambda$ should be considered as a free 
parameter. In the numerical calculation we have used the value 
$\lambda = S/L = 0.25$ that leads to the best agreement with the data 
\cite{abe}. 

The calculated inclusive densities of different secondaries at RHIC, 
$\sqrt{s} = 200.$ GeV, and LHC, $\sqrt{s} = 14.$ TeV, energies are presented in 
Table 1, where one can see that the agreement of the QGSM calculations with 
RHIC experimental data \cite{abe} is reasonably good. 

\begin{table}[t]
\caption{The QGSM results for midrapidity yields $dn/dy$ ($\vert y \vert < 0.5$) for
different secondaries at RHIC}
\vspace{-0.125cm} 
{\footnotesize \hspace{0.75cm}and LHC energies. The results for
$\varepsilon = 0.024$ are presented only when different from the case
$\varepsilon = 0$.}
\begin{center}
\vskip 8pt
\begin{tabular}{|c|r|r|r|r|r|} \hline
& \multicolumn{3}{c|}{RHIC ($\sqrt{s} = 200.$ GeV)} & 
\multicolumn{2}{c|}{LHC ($\sqrt{s} = 14.$ TeV)} \\ \cline{2-6}
Particle & $\varepsilon = 0$ & $\varepsilon = 0.024$ & STAR Collaboration \cite{abe} &
$\varepsilon = 0$ & $\varepsilon = 0.024$ \\   \hline
$\pi^+$         & 1.27    &        &                     & 2.54  & \\ 
$\pi^-$         & 1.25    &        &                     & 2.54  & \\
$K^+$           & 0.13    &        & $0.14 \pm 0.01$     & 0.25  &  \\
$K^-$           & 0.12    &        & $0.14 \pm 0.01$     & 0.25  &  \\
$p$             & 0.0755  & 0.0861 &                     & 0.177 & 0.184  \\
$\overline{p}$  & 0.0707  &        &                     & 0.177 &        \\
$\Lambda$       & 0.0328  & 0.0381 & $0.0385 \pm 0.0035$ & 0.087 & 0.0906 \\
$\overline{\Lambda}$ & 0.0304  &   & $0.0351 \pm 0.0032$ & 0.0867 &  \\
$\Xi^-$      & 0.00306  & 0.00359 & $0.0026 \pm 0.0009$ & 0.0108 & 0.0112 \\
$\overline{\Xi}^+$ & 0.00298 &     & $0.0029 \pm 0.001$  & 0.0108&  \\
$\Omega^-$       & 0.00020  & 0.00025 & * & 0.000902 & 0.000934 \\
$\overline{\Omega}^+$ & 0.00020  &    & * & 0.000902 &  \\
\hline
\end{tabular}
\end{center}
\hspace{1.2cm}$^* dn/dy(\Omega^- + \overline{\Omega}^+) = 0.00034 \pm 0.00019$
\end{table}

\begin{figure}[h]
\centering
\vskip -0.8cm
\includegraphics[width=.75\hsize]{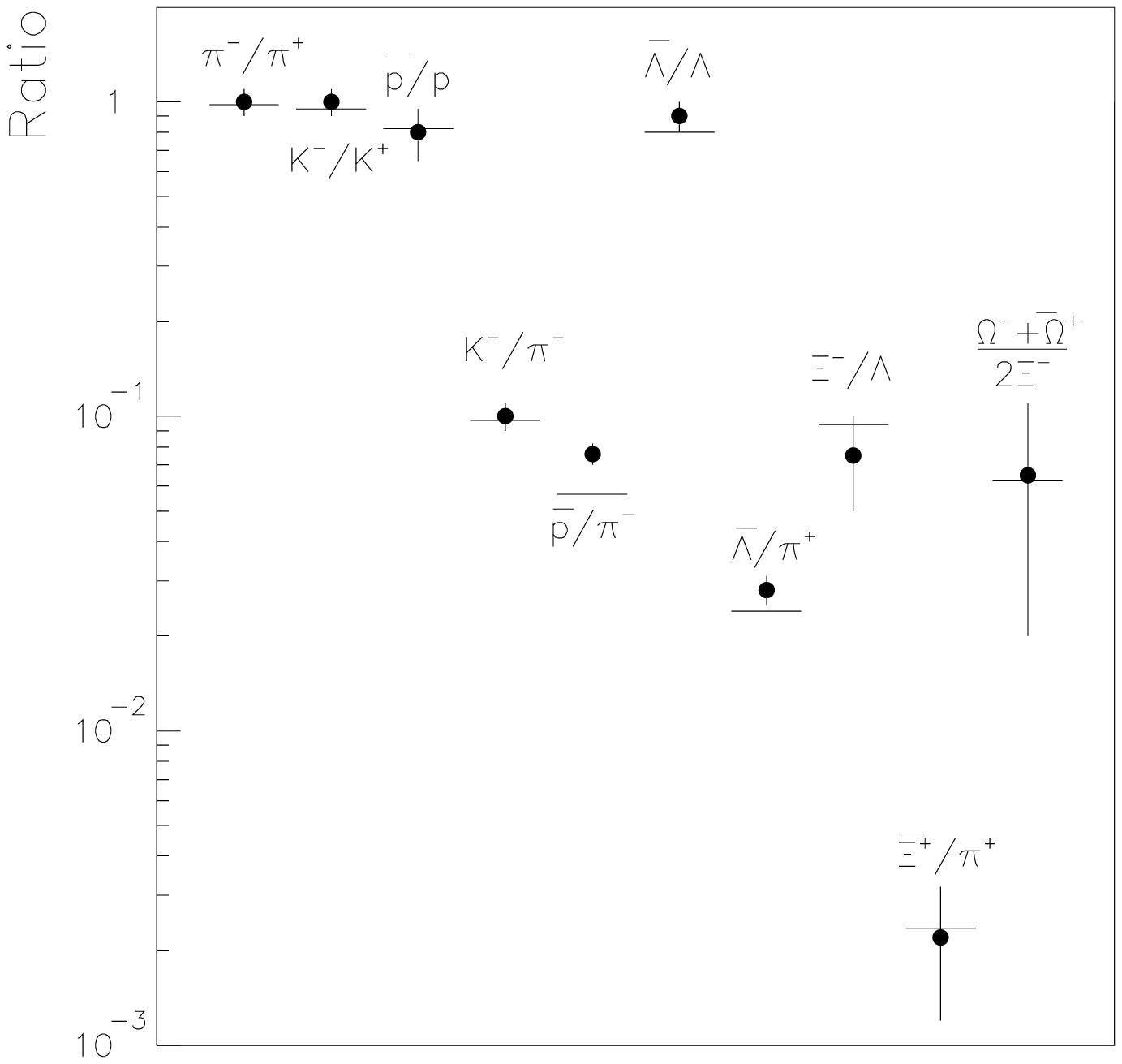}
\vskip -1.7cm
%%%%\caption{\footnotesize
%%%%Ratios of different secondaries produced in midrapidity region in $pp$ 
%%%%collisions at $\sqrt{s}=200.$GeV. Short horizontal solid lines show the results 
%%%%of the QGSM calculations.}
\caption{\footnotesize
Ratios of different secondaries produced in midrapidity region in $pp$
collisions at $\sqrt{s}=200.$GeV.}
\vspace{-0.125cm}
{\footnotesize
\hspace{-5.cm}Short horizontal solid lines show the results
of the QGSM calculations.}
\vskip -0.3cm
\label{Fig2}
\end{figure}

In Fig.~\ref{Fig2} we reproduce the experimental data on ratios of yields of different 
secondaries~\cite{abe} together with our calculations. Agreement is good 
except for the point of the $\bar{p}/\pi^-$ ratio. From the comparison 
of our results with experimental data presented in Table~1 and Fig.~\ref{Fig2}
we can conclude that the universal parameter $\lambda$ = 0.25 describes the ratios of
$\Lambda/p$, $\Xi/\Lambda$, and $\Omega/\Xi$ production in a reasonable way.

\section*{Acknowledgments}
This paper was supported by Ministerio de Educaci\'on y Ciencia of Spain under 
the Spanish Consolider-Ingenio 2010 Programme CPAN (CSD2007-00042) and project 
FPA 2005--01963, by Xunta de Galicia and, in part, by grants 
RFBR-07-02-00023 and RSGSS-1124.2003.2.

\end{document}